\pdfoutput=1
\documentclass{article}
\usepackage{spconf,amsmath,graphicx}
\usepackage{iftex}
\ifxetex
  \usepackage{fontspec}
\fi
\usepackage{booktabs}
\usepackage{url}
\usepackage{pgfplots}
\pgfplotsset{compat=1.18}
\usepackage{tikz}
\usetikzlibrary{arrows.meta}
\newcommand{\olkVCer}{26.64}
\newcommand{\olkVWer}{47.89}
\newcommand{\olkVParams}{34M}
\newcommand{\olkSpeakersPaper}{1,107}
\newcommand{\corpusSpeakers}{1,000}
\newcommand{\corpusRecordings}{12,000}
\newcommand{\corpusUniqueHours}{727}
\newcommand{\tlSpeakers}{893}
\newcommand{\vlSpeakers}{107}
\newcommand{\storePx}{160}
\newcommand{\storeFps}{25}
\newcommand{\inputPx}{96}
\newcommand{\mouthPx}{57}
\newcommand{\vlUttSeenShare}{63.2\%}
\newcommand{\seenShareCellFamOne}{60.3\%}
\newcommand{\seenShareCellFamTwo}{55.4\%}
\newcommand{\cellSize}{1,800}
\newcommand{\nBoot}{2,000}
\newcommand{\unseenMedianSec}{5.71}
\newcommand{\seenMedianSec}{4.95}
\newcommand{\encParams}{175.0M}
\newcommand{\decParams}{56.8M}
\newcommand{\totalParams}{232M}
\newcommand{\dModel}{768}
\newcommand{\encBlocks}{12}
\newcommand{\decBlocks}{6}
\newcommand{\vocabCtc}{70}
\newcommand{\ctcW}{1.0}
\newcommand{\attW}{0.3}
\newcommand{\lrPeak}{$2\times10^{-4}$}
\newcommand{\lrMin}{$2\times10^{-5}$}
\newcommand{\decaySteps}{800k}
\newcommand{\framesPerStep}{9,600}
\newcommand{\timeMaskAfter}{20k}
\newcommand{\jitterPx}{4}
\newcommand{\beamSize}{5}
\newcommand{\ctcwJoint}{0.2}
\newcommand{\eTwoUtts}{2.12M}
\newcommand{\eThreeUtts}{434k}
\newcommand{\gpuName}{RTX PRO 6000 Blackwell (96 GB)}
\newcommand{\gpuHoursEtwo}{140}
\newcommand{\cerEtwoBdfhA}{9.95}
\newcommand{\ciEtwoBdfhA}{[9.18, 10.72]}
\newcommand{\werEtwoBdfhA}{20.47}
\newcommand{\cerEtwoCegiA}{12.19}
\newcommand{\ciEtwoCegiA}{[11.35, 13.16]}
\newcommand{\werEtwoCegiA}{24.08}
\newcommand{\exactEtwoBdfhA}{48.8\%}
\newcommand{\exactEtwoUnseenBdfhA}{9.7\%}
\newcommand{\nEtwoBdfhA}{1,790}
\newcommand{\nEtwoCegiA}{1,792}
\newcommand{\marginCerLo}{14.5}
\newcommand{\marginCerHi}{16.7}
\newcommand{\cerGreedyEtwoBdfhA}{14.06}
\newcommand{\werGreedyEtwoBdfhA}{29.60}
\newcommand{\cerGreedyEtwoCegiA}{16.82}
\newcommand{\werGreedyEtwoCegiA}{34.13}
\newcommand{\gapEtwoVlC}{2.03}
\newcommand{\gapEtwoNewI}{6.51}
\newcommand{\bandUpperEtwoNew}{5.77}
\newcommand{\bandLevelEtwoNew}{1.28}
\newcommand{\bandLowerEtwoNew}{0.36}
\newcommand{\upperPenaltyAbout}{six}
\newcommand{\cerUnseenJointBdfh}{19.00}
\newcommand{\ciUnseenJointBdfh}{[18.21, 19.78]}
\newcommand{\werUnseenJointBdfh}{36.87}
\newcommand{\cerUnseenJointCegi}{21.52}
\newcommand{\ciUnseenJointCegi}{[20.67, 22.33]}
\newcommand{\werUnseenJointCegi}{40.57}
\newcommand{\expertShareSeenLo}{29}
\newcommand{\expertShareSeenHi}{36}
\newcommand{\expertShareUnseenLo}{60}
\newcommand{\expertShareUnseenHi}{66}
\newcommand{\censusSpeakers}{106}
\newcommand{\censusUtts}{11,835}
\newcommand{\censusSpkMin}{1.0}
\newcommand{\censusSpkMax}{52.2}
\newcommand{\censusSpkMedian}{7.3}
\newcommand{\baselineBeam}{1}
\newcommand{\bandUpperEthree}{18.93}
\newcommand{\bandLevelEthree}{4.52}
\newcommand{\bandLowerEthree}{4.79}
\newcommand{\cerEthreeJointSeenBdfh}{9.92}
\newcommand{\werEthreeJointSeenBdfh}{21.08}
\newcommand{\werEthreeJointSeenCegi}{24.60}
\newcommand{\werEthreeJointUnseenBdfh}{36.69}
\newcommand{\werEthreeJointUnseenCegi}{40.14}
\newcommand{\cerEthreeJointSeenCegi}{12.25}
\newcommand{\cerEthreeJointUnseenBdfh}{18.24}
\newcommand{\cerEthreeJointUnseenCegi}{20.97}
\newcommand{\trajBaseHi}{50.9}
\newcommand{\trajBaseLo}{26.1}
\newcommand{\trajPenHMean}{5.2}
\newcommand{\trajPenIMean}{7.0}
\newcommand{\jointGainEndLo}{5.4}
\newcommand{\jointGainEndHi}{5.9}
\newcommand{\eTwoHours}{3,166}
\newcommand{\eTwoSpeakers}{876}
\newcommand{\cerSeenBdfh}{9.95}
\newcommand{\cerSeenCegi}{12.19}
\newcommand{\cerUnseenBdfh}{19.00}
\newcommand{\cerUnseenCegi}{21.52}
\newcommand{\expertShareCorpus}{30\%}
\newcommand{\expertSpontShare}{half}
\newcommand{\spontRuleThresh}{20\%}
\newcommand{\spontRuleExpertMedian}{50\%}
\newcommand{\censusCer}{14.65}
\newcommand{\censusSpkQone}{2.9}
\newcommand{\censusSpkQthree}{17.3}
\newcommand{\censusOrdinary}{6.17}
\newcommand{\ciCensusOrdinary}{[6.0, 6.4]}
\newcommand{\nOrdinarySpk}{80}
\newcommand{\censusExpert}{31.26}
\newcommand{\ciCensusExpert}{[30.7, 31.9]}
\newcommand{\nExpertSpk}{26}
\newcommand{\censusF}{14.54}
\newcommand{\censusM}{14.75}
\newcommand{\cOrdReadSeen}{4.79}
\newcommand{\ciOrdReadSeen}{[4.6, 5.0]}
\newcommand{\cOrdReadUnseen}{11.82}
\newcommand{\ciOrdReadUnseen}{[11.3, 12.3]}
\newcommand{\cExpReadSeen}{13.31}
\newcommand{\ciExpReadSeen}{[12.1, 14.5]}
\newcommand{\cExpReadUnseen}{22.32}
\newcommand{\ciExpReadUnseen}{[21.0, 23.8]}
\newcommand{\cExpSpont}{35.03}
\newcommand{\ciExpSpont}{[34.5, 35.7]}
\newcommand{\nOrdReadSeen}{6,561}
\newcommand{\nOrdReadUnseen}{1,463}
\newcommand{\nExpReadSeen}{474}
\newcommand{\nExpReadUnseen}{407}
\newcommand{\nExpSpont}{2,579}
\newcommand{\effWordingOrd}{7.0}
\newcommand{\effWordingExp}{9.0}
\newcommand{\effDeliverySeen}{8.5}
\newcommand{\effDeliveryUnseen}{10.5}
\newcommand{\effSpont}{12.7}
\newcommand{\evalUttsLo}{85}
\newcommand{\evalUttsHi}{203}
\newcommand{\budgetsChunks}{1, 2, 4, 8}
\newcommand{\budgetsMin}{4, 7, 14, 29}
\newcommand{\lrFT}{$2\times10^{-5}$}
\newcommand{\lrLoRA}{$2\times10^{-4}$}
\newcommand{\ftSteps}{200}
\newcommand{\ftFrames}{1,200}
\newcommand{\valEvery}{20}
\newcommand{\forgetPopN}{1,786}
\newcommand{\forgetPopSpk}{50}
\newcommand{\loraDecLinears}{48}
\newcommand{\gainLoraRthirtytwo}{2.96}
\newcommand{\gainLoraRthirtytwoSeven}{1.60}
\newcommand{\loraRankBest}{32}
\newcommand{\loraTrainableBest}{10.9M}
\newcommand{\loraShareBest}{4.6\%}
\newcommand{\loraFileMBBest}{44}
\newcommand{\pilotTwelve}{twelve}
\newcommand{\gainFTtwelveTwentyNine}{4.22}
\newcommand{\driftFTtwelve}{8.27}
\newcommand{\gainRtwelveSeven}{2.13}
\newcommand{\gainRtwelveTwentyNine}{3.58}
\newcommand{\driftRtwelve}{0.98}
\newcommand{\loraShareOfFTtwelve}{85\%}
\newcommand{\loraShareOfForgettingTwelve}{12\%}
\newcommand{\relRecoverOrdLoR}{11\%}
\newcommand{\relRecoverOrdHiR}{46\%}
\newcommand{\relRecoverExpLoR}{7\%}
\newcommand{\relRecoverExpHiR}{10\%}
\newcommand{\randomRelMedianR}{16\%}
\newcommand{\gainRsixFour}{1.32}
\newcommand{\gainRsixFourteen}{2.11}
\newcommand{\driftRmax}{2.50}
\newcommand{\coldRangeTwelveLo}{8.4}
\newcommand{\coldRangeTwelveHi}{43.8}
\newcommand{\pthreeRUpper}{4.09}
\newcommand{\pthreeRLevel}{3.37}
\newcommand{\pthreeRLower}{3.45}
\newcommand{\pthreeRCells}{68}
\newcommand{\pthreeRExcluding}{60}
\newcommand{\bOneBothOrd}{2.07}
\newcommand{\bOneEncOrd}{2.09}
\newcommand{\bOneDecOrd}{0.09}
\newcommand{\bOneBothProf}{2.33}
\newcommand{\bOneEncProf}{2.57}
\newcommand{\bOneDecProf}{0.49}
\newcommand{\randomN}{eight}
\newcommand{\randomGainFT}{2.42}
\newcommand{\randomGainR}{1.96}
\newcommand{\censusSpearman}{0.989}
\newcommand{\censusTopTen}{nine}
\newcommand{\jOrdReadSeen}{1.56}
\newcommand{\jOrdReadUnseen}{6.41}
\newcommand{\jExpReadSeen}{6.88}
\newcommand{\jExpReadUnseen}{15.32}
\newcommand{\jExpSpont}{29.10}
\newcommand{\jEffWordingOrd}{4.9}
\newcommand{\jEffWordingExp}{8.4}
\newcommand{\jEffSpont}{13.8}
\newcommand{\bNineHours}{587}
\newcommand{\Mnine}{M$_9$}
\newcommand{\Mone}{M$_1$}
\newcommand{\seedGapLo}{1.7}
\newcommand{\seedGapHi}{2.5}
\newcommand{\seedStep}{700k}
\newcommand{\jointSlowdown}{25 - 30}
\newcommand{\loraEncLinearsAll}{72}
\newcommand{\unseenTierN}{3,578}
\newcommand{\lpConformerWer}{12.8\% WER}
\newcommand{\vallrWer}{18.7\% WER}
\newcommand{\jointGainSeenLo}{4.1}
\newcommand{\jointGainSeenHi}{4.6}
\newcommand{\jointCaseGainLo}{3}
\newcommand{\jointCaseGainHi}{7}

\title{Personalized Korean Lipreading as Visual Speech Recognition:\\
Transfer, Census and Adaptation on OLKAVS}
\name{Se Un Park$^{\ast}$, Hakjun Kim, Taehoon Roh, Junyoung Park$^{\ast}$\thanks{$^{\ast}$Corresponding authors.}}
\address{UX Factory, Inc.}

\begin{document}
\ninept
\maketitle
\begin{abstract}
We present a personalized Korean visual speech recognition (VSR) system
and quantify, on the nine-camera OLKAVS corpus, the gap between the
population-level benchmark score and an individual user's error. A
video-only Conformer initialized from English-trained weights attains
\cerSeenBdfh{} - \cerSeenCegi{}\% character error rate (CER) under the
corpus protocol against the published \olkVCer{}, and \cerUnseenBdfh{} -
\cerUnseenCegi{} on unseen wording. Per speaker, CER spans \censusSpkMin{}
to \censusSpkMax{}\%, with seen wording lowering CER by \effWordingOrd{} -
\effWordingExp{} points and professional delivery and spontaneous speech
raising it by \effDeliverySeen{} - \effDeliveryUnseen{} and \effSpont{}
points. A
low-rank adapter with \loraShareBest{} of the parameters, trained on 4 to
29 minutes of the user's frontal video, lowers the CER of \pilotTwelve{}
high-error speakers by \gainRtwelveSeven{} to \gainRtwelveTwentyNine{}
points, transfers to every camera without loss, and keeps
\loraShareOfFTtwelve{} of the full fine-tuning gain at
\loraShareOfForgettingTwelve{} of its cost to other speakers. Cameras above
the mouth plane add about \upperPenaltyAbout{} CER points as a constant
offset that training on all views keeps small.
\end{abstract}

\section{Introduction}
\label{sec:intro}

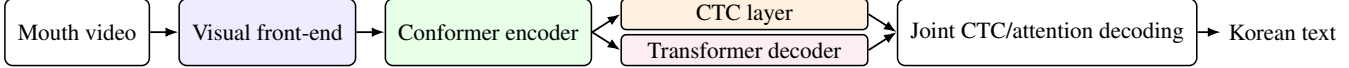
\begin{figure*}[t]
  \centering
  \resizebox{\textwidth}{!}{\begin{tikzpicture}[x=1pt, y=1pt, font=\small,
  box/.style={draw, rounded corners=3pt, align=center, inner xsep=5pt, inner ysep=4pt, minimum height=26pt, line width=0.6pt},
  thin/.style={draw, rounded corners=2pt, align=center, inner xsep=5pt, inner ysep=1.5pt, minimum height=12pt, line width=0.6pt},
  fe/.style={box, fill=blue!7}, enc/.style={box, fill=green!9},
  ctc/.style={thin, fill=orange!12}, dec/.style={thin, fill=purple!7},
  ar/.style={-{Latex[length=5pt]}, line width=0.7pt},
  note/.style={font=\small, align=center}]
  \node[box] (in) at (0,0) {Mouth video};
  \node[fe, anchor=west] (fe) at ([xshift=11pt]in.east) {Visual front-end};
  \node[enc, anchor=west] (enc) at ([xshift=11pt]fe.east) {Conformer encoder};
  \node[ctc, anchor=west, minimum width=96pt] (ctc) at ([xshift=11pt, yshift=7pt]enc.east) {CTC layer};
  \node[dec, anchor=west, minimum width=96pt] (dec) at ([xshift=11pt, yshift=-7pt]enc.east) {Transformer decoder};
  \node[box, anchor=west] (joint) at ([xshift=11pt]ctc.east |- enc.east) {Joint CTC/attention decoding};
  \node[note, anchor=west] (out) at ([xshift=9pt]joint.east) {Korean text};
  \draw[ar] (in) -- (fe);
  \draw[ar] (fe) -- (enc);
  \draw[ar] (enc.east) -- (ctc.west);
  \draw[ar] (enc.east) -- (dec.west);
  \draw[ar] (ctc.east) -- (joint.west);
  \draw[ar] (dec.east) -- (joint.west);
  \draw[ar] (joint) -- (out);
\end{tikzpicture}}
  \par\vspace{-6pt}
  \caption{Our VSR architecture. Mouth video (\inputPx{} pixels at
  \storeFps{} fps) passes through the visual front-end (3D convolution and
  ResNet-18) and the Conformer encoder (\encBlocks{} blocks of width
  \dModel{}). A CTC layer over \vocabCtc{} jamo symbols (loss weight \ctcW{})
  and a \decBlocks{}-block Transformer decoder (label-smoothed
  cross-entropy, weight \attW{}) share the encoder output, and joint
  CTC/attention decoding yields Korean text.}
  \label{fig:arch}
\end{figure*}

Lipreading could let a conversational assistant understand its user where
a microphone fails, on a crowded train or in a noisy street, and let a user
in a quiet office address a coding assistant by mouthing words with little
or no sound. In these settings the system serves one person, who holds a phone camera
at an arbitrary angle and speaks unscripted sentences. A VSR benchmark, by contrast, reports a single score over a population of
speakers, mostly filmed frontally and often reading sentences drawn from
a pool shared with the training data. We quantify the gap between the two on OLKAVS \cite{park2024olkavs}, a
\olkSpeakersPaper{}-speaker Korean audio-visual corpus, with a model that
outperforms the published baseline by a wide margin, and we measure what a
few minutes of the user's own video and the camera angle change.

Three factors separate the benchmark score from the individual user's
error. Speakers differ far more in appearance than in voice, so a
population mean hides a spread that matters to each user. Previous speaker
adaptation of VSR models, by user-dependent padding \cite{kim2022padding}
and by vision-and-language prompting \cite{yeo2024personalized}, reports
gains on English data. Here we measure the gain as a function of the
minutes of the user's video available, its transfer across camera views,
and its cost for other speakers, for full fine-tuning and for a low-rank
adapter \cite{hu2021lora}.
Scripted corpora share sentences across partitions, so part of a reported
score reflects memorized wording rather than lipreading
\cite{kapoor2023leakage,tseng2025contamination,jain2026lipreadinggap}. We separate the wording effect from speaker type and speech mode, the
latter a distinct problem for spontaneous Korean even in audio-based
recognition \cite{bang2020ksponspeech}. Finally, benchmarks are frontal and level whereas hand-held cameras may
not be. Multi-view VSR has concentrated on horizontal camera
placements and yaw-invariant modeling
\cite{anina2015ouluvs2,chung2017profile,koumparoulis2018view2view,petridis2017multiview,cheng2020poseinvariant},
whereas elevation, the angle that a hand-held camera varies most, has been
measured only on a closed vocabulary \cite{isobe2022icpram} or on
unmatched content \cite{park2024olkavs}, and studies of human
speechreading do not predict an asymmetric loss for elevated views
\cite{erber1974angle,jordan2011halfface,preminger1998masking}.

We address all three with a single model, a \totalParams{}-parameter hybrid
CTC/attention Conformer initialized from an English-trained VSR model
\cite{ma2023autoavsr} and trained on video from the OLKAVS training
partition only. Our contributions are as follows.
(i) A new state of the art for Korean sentence-level VSR under the corpus
protocol, \cerSeenBdfh{} - \cerSeenCegi{}\% CER against the published
\olkVCer{} \cite{park2024olkavs,olkavs2023github} (Sec.~\ref{ssec:official}).
(ii) The error distribution over every held-out speaker rather than its
mean, and a decomposition of the official score into wording, delivery and
spontaneity with a speech mode recovered from sentence sharing
(Sec.~\ref{ssec:census}).
(iii) Personalization measured as a function of minutes of the user's own
video on speakers with high base error, with all evaluation sentences
held out. One frontal recording transfers to every camera, a
low-rank adapter retains most of the full fine-tuning gain at a fraction
of its cost to other speakers, and the gain is confined to the visual encoder (Sec.~\ref{ssec:personal}).
(iv) A paired evaluation of camera elevation on identical utterances shows
that
cameras above the mouth plane add about \upperPenaltyAbout{} CER points,
additively throughout training, and that training on all views keeps the
penalty at that level (Sec.~\ref{ssec:elevation}).

\section{Corpus, model and protocol}
\label{sec:method}

\subsection{Corpus}
\label{ssec:corpus}

OLKAVS \cite{park2024olkavs,aihub538} comprises \corpusSpeakers{} speakers
in \corpusRecordings{} five-minute studio recordings, each captured
simultaneously by five of nine cameras, for about \corpusUniqueHours{}
hours of unique transcribed speech. The official partition is
speaker-disjoint (\tlSpeakers{} training and \vlSpeakers{} validation
speakers). Every model in this paper is trained on the training partition
only and evaluated on validation speakers never seen in training. We store
one \storePx{}-pixel grayscale crop of the lower face per recording at
\storeFps{}~fps and feed the model a \inputPx{}-pixel center crop
(mouth width about \mouthPx{} pixels). The cameras form three
rows.\footnote{The camera array is documented without elevation or azimuth
angles.} A is frontal, B, I and H are above it, C and G are level with it,
and D, E and F are below. Each recording uses camera A together with either
\{B,D,F,H\} (camera group~1) or \{C,E,G,I\} (camera group~2). The two groups
therefore partition the validation set, and a view is compared only with
camera A of its own group on identical utterances. CERs are never pooled
across groups.

\textbf{Speaker type and speech mode.} The corpus annotates
\expertShareCorpus{} of the speakers as speech professionals (announcers, actors and trainees, denoted P, against O for ordinary
speakers). Roughly
\expertSpontShare{} of their recordings are spontaneous (impromptu) speech
on a news article, whereas ordinary speakers only read scripts
\cite{park2024olkavs}. No per-recording annotation distinguishes script
reading from spontaneous speech, which we call the speech mode. Scripts are
drawn from a shared pool and recur across speakers, whereas a spontaneous
sentence cannot, so we label a recording as spontaneous when fewer than
\spontRuleThresh{} of its sentences are spoken by any other speaker. The
rule marks a median \spontRuleExpertMedian{} of each professional's
recordings and none of any ordinary speaker's, in agreement with the
proportions reported in \cite{park2024olkavs}. The model operates on jamo,
the letters of written Korean and roughly its phonemic units (19 initial
consonants, 21 vowels and 27 final consonants). Two or three jamo compose
one syllable, so a jamo sequence maps to text without a lexicon, and CER is
computed over composed characters, as in the published protocol.

\subsection{Model and training}
\label{ssec:model}

We present the model architecture in Fig.~\ref{fig:arch}. The encoder is the Auto-AVSR visual
front-end and Conformer back-end \cite{ma2021avsrconformer,ma2023autoavsr,gulati2020conformer}
with rotary position encoding \cite{su2024roformer}, operating on
\inputPx{}-pixel frames at the full \storeFps{}~Hz frame rate, and a linear
CTC layer and a \decBlocks{}-layer attention decoder share a vocabulary of
\vocabCtc{} positional jamo symbols. The model has \encParams{} encoder and
\decParams{} decoder parameters, about \totalParams{} in total, and no
audio branch. The encoder and decoder are initialized from the LRS3
visual-only Auto-AVSR checkpoint \cite{ma2023autoavsr}, trained on English
video with automatically generated transcripts. The vocabulary-dependent
layers and the CTC layer are randomly initialized, and all parameters are
trained. Training uses the hybrid CTC/attention loss with the loss weights given in
Fig.~\ref{fig:arch} \cite{watanabe2017hybrid}. No camera, view, speaker or speech-mode information is
supplied to the model in training or evaluation, so any view robustness of
an all-view model is learned from the data mixture alone. We train with AdamW to the end of a cosine schedule without early
stopping.\footnote{Peak learning rate \lrPeak{}, cosine decay to \lrMin{}
over \decaySteps{} steps, \framesPerStep{} frames per step, crop jitter of
$\pm$\jitterPx{} pixels, horizontal flipping, and time masking of up to 10\%
of the frames of a clip after step \timeMaskAfter{}. The validation curve
was flat over the final 200k steps for both models.} \Mnine{} is trained on all nine
views of the training partition (\eTwoUtts{} utterances, \eTwoHours{} hours
of video, \eTwoSpeakers{} speakers) and is the model behind every result below unless stated otherwise.
\Mone{} is its counterpart trained on frontal views only (\eThreeUtts{}
utterances), differing only in that training condition. Training \Mnine{} took \gpuHoursEtwo{}
GPU-hours on one \gpuName{}. Reported scores use joint CTC/attention beam
search \cite{watanabe2017hybrid} (beam \beamSize{}, CTC weight
\ctcwJoint{}), as our deployment configuration. Only the per-view analyses, which require many evaluations, use greedy CTC
decoding, which needs no decoder, costs \jointSlowdown{} times less, and
preserves the ranking of the views.

\subsection{Evaluation protocol}
\label{ssec:protocol}

We implement the published protocol \cite{park2024olkavs,olkavs2023github};
it trains on the training partition and scores camera A of the separate
validation speakers with hybrid decoding and a CER pooled over utterances,
spaces included. Every evaluation
reads a fixed list of utterance identifiers, and for the official score we
draw \cellSize{} utterances per view from the five-camera-matched
validation utterances of each camera group. The two partitions are drawn
from one sentence pool, and \vlUttSeenShare{} of the validation utterances
are sentences that also occur, spoken by other people, in the training
partition. We therefore define two subsets per camera group. The
\emph{seen-wording subset} is drawn from the full protocol population and
corresponds to what the published protocol measures
(\seenShareCellFamOne{} / \seenShareCellFamTwo{} of its sentences in camera
group 1 / group 2 occur in the training text). The \emph{unseen-wording subset} excludes every
sentence whose text occurs in the training transcripts (after text
normalization, none of its \unseenTierN{} sentences does). Because unseen sentences are longer (median \unseenMedianSec{} against
\seenMedianSec{}~s), the seen subset is sampled to match their duration
histogram. Every evaluation is a single pass with a per-utterance record,
so every difference between cameras, subsets or models is computed on
identical utterances, with 95\% percentile confidence intervals (CI) from
\nBoot{} bootstrap resamples of the utterances \cite{bisani2004bootstrap},
each difference evaluated on the same resamples for both sides. A view gap is the CER of
a view minus that of camera A of its own group.

\subsection{Speaker-level evaluation and personalization protocol}
\label{ssec:personalprotocol}

For every validation speaker with all twelve recordings, three recordings
are held out as an evaluation set (\evalUttsLo{} - \evalUttsHi{}
utterances each), and the base model, \Mnine{} before any speaker
adaptation, scores all of them (\censusSpeakers{} speakers, \censusUtts{}
utterances, camera A). We call this evaluation the census and assign each utterance to a case by speaker type, speech mode and whether its sentence
occurs in the training text. For personalization the same three recordings
serve as the evaluation set at every budget, one further recording is held out to score the training checkpoints and
select the best step, and the remaining recordings provide nested budgets
of \budgetsChunks{} recordings, about \budgetsMin{} minutes of frontal
video. Recordings are the unit of the split, and any sentence shared
between personalization and evaluation recordings is removed. The
reference population for the effect on other speakers is the frontal
unseen-wording subset (\forgetPopN{} utterances of \forgetPopSpk{} other
validation speakers). The \pilotTwelve{} pilot speakers were selected from the census for high
base-model CER, six P and six O spanning \coldRangeTwelveLo{} - \coldRangeTwelveHi{}\%,
and \randomN{} further speakers drawn uniformly from the census serve as
a check on the selection.

Every configuration resumes the trained model, runs \ftSteps{} steps on
the personalization recordings, and scores the checkpoint with the best
validation CER on the held-out recording.\footnote{\ftFrames{} frames per step at the final
learning rate of the schedule (\lrFT{} for FT and FT-FE, \lrLoRA{} for the
adapter), training augmentations retained, validation every \valEvery{}
steps. The adapter also freezes the front-end's batch-normalization
statistics and sits beside the \loraEncLinearsAll{} attention and
feed-forward linear layers of the Conformer and the \loraDecLinears{} of
the decoder.} Full fine-tuning (FT) updates all parameters, FT with the front-end
frozen (FT-FE) keeps the 3D stem and ResNet trunk fixed, and
LoRA-\loraRankBest{}
\cite{hu2021lora} freezes the base model and trains a rank-\loraRankBest{}
adapter ($\alpha=64$, \loraTrainableBest{} parameters, \loraShareBest{})
in parallel with every attention and feed-forward linear layer of the
Conformer and the decoder, and the adapter is merged into the base weights
for scoring.

\section{Results}
\label{sec:results}

All differences are in CER points. The personalization gain is the base
CER minus the adapted CER on the speaker's own evaluation set, and the
degradation of other speakers is the adapted minus the base CER on the
reference population, so a positive value is an improvement in the first
case and a loss in the second.

\subsection{Official protocol}
\label{ssec:official}

Table~\ref{tab:official} compares \Mnine{} at the end of its schedule with
the published baseline. Under the published protocol (Sec.~\ref{ssec:protocol}) it attains
\cerEtwoBdfhA{} (95\% CI \ciEtwoBdfhA{}) and \cerEtwoCegiA{}
\ciEtwoCegiA{}\% CER on the two camera groups, \marginCerLo{} -
\marginCerHi{} points below the published V-model, with less than half of
its word error rate and \exactEtwoBdfhA{} of the utterances recognized
exactly.\footnote{We verified the comparison against the reference
implementation, with identical training partition, validation speakers,
camera, metric and hybrid decoding. Our rows are \cellSize{}-utterance
samples per camera group, and both systems rely on English pretraining
(the V-model through a frozen front-end, ours through its
initialization), so \totalParams{} against \olkVParams{} parameters is not
an efficiency comparison.} The margin does not depend on the decoding method. Greedy CTC decoding attains \cerGreedyEtwoBdfhA{} /
\cerGreedyEtwoCegiA{}, and joint decoding improves on it by
\jointGainSeenLo{} - \jointGainSeenHi{} points on the seen-wording subset
and \jointGainEndLo{} - \jointGainEndHi{} on the unseen-wording
subset.\footnote{A second seed of \Mnine{}, scored at step \seedStep{} on
the same unseen-wording sets, differs from the first by \seedGapLo{} -
\seedGapHi{} points, the magnitude of run-to-run variation.} On the
unseen-wording subset the same model attains \cerUnseenJointBdfh{}
\ciUnseenJointBdfh{} and \cerUnseenJointCegi{} \ciUnseenJointCegi{}\%,
still below the published seen-subset figure, while exact matches fall to
\exactEtwoUnseenBdfhA{}. The seen subset therefore measures memorization to a large extent, and the
unseen subset measures lipreading. The two subsets are matched in length but not in speaker composition
(professionals speak \expertShareSeenLo{} - \expertShareSeenHi{}\% of the
seen and \expertShareUnseenLo{} - \expertShareUnseenHi{}\% of the unseen
utterances, since their spontaneous sentences cannot occur in the training
text), so their difference combines a wording effect with a change of
population, which the census separates.

\begin{table}[t]
  \centering
  \caption{Official protocol on camera A, camera group 1/group 2, CER and
  WER (\%). Parentheses give the training views. Published row as released
  \cite{olkavs2023github} (hybrid decoding, beam \baselineBeam{}, parameter
  count without its frozen front-end). joint = CTC/attention, greedy =
  greedy CTC, unseen = unseen-wording subset. \nEtwoBdfhA{} /
  \nEtwoCegiA{} utterances per row.}
  \label{tab:official}
  \setlength{\tabcolsep}{2.5pt}
  \begin{tabular}{@{}llrr@{}}
    \toprule
    System (views, decoding) & Params & CER & WER \\
    \midrule
    V-model~\cite{park2024olkavs} (not stated) & \olkVParams{} & \olkVCer{} & \olkVWer{} \\
    \midrule
    \Mnine{} (all views, joint) & \totalParams{} & \cerEtwoBdfhA{}/\cerEtwoCegiA{} & \werEtwoBdfhA{}/\werEtwoCegiA{} \\
    \Mnine{} (all views, greedy) & & \cerGreedyEtwoBdfhA{}/\cerGreedyEtwoCegiA{} & \werGreedyEtwoBdfhA{}/\werGreedyEtwoCegiA{} \\
    \Mnine{}, unseen (joint) & & \cerUnseenJointBdfh{}/\cerUnseenJointCegi{} & \werUnseenJointBdfh{}/\werUnseenJointCegi{} \\
    \midrule
    \Mone{} (frontal only, joint) & \totalParams{} & \cerEthreeJointSeenBdfh{}/\cerEthreeJointSeenCegi{} & \werEthreeJointSeenBdfh{}/\werEthreeJointSeenCegi{} \\
    \Mone{}, unseen (joint) & & \cerEthreeJointUnseenBdfh{}/\cerEthreeJointUnseenCegi{} & \werEthreeJointUnseenBdfh{}/\werEthreeJointUnseenCegi{} \\
    \bottomrule
  \end{tabular}
\end{table}

\subsection{Per-speaker error distribution}
\label{ssec:census}

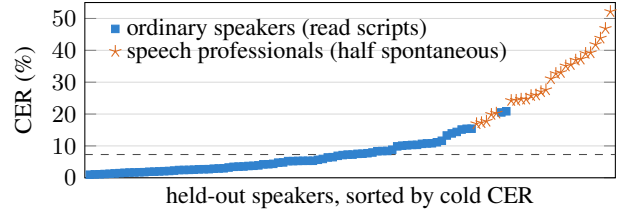
\begin{figure}[t]
  \centering
  \begin{tikzpicture}
\definecolor{cOrd}{RGB}{49,130,206}
\definecolor{cExp}{RGB}{221,107,32}
\begin{axis}[
  width=\columnwidth, height=3.9cm,
  xmin=0, xmax=107, ymin=0, ymax=55,
  xtick=\empty, ytick={0,10,20,30,40,50},
  xlabel={held-out speakers, sorted by cold CER}, ylabel={CER (\%)},
  label style={font=\small}, tick label style={font=\small},
  xlabel near ticks, ylabel near ticks,
  ymajorgrids, grid style={gray!30, very thin},
  legend style={font=\small, at={(0.03,0.97)}, anchor=north west, draw=none, fill=none, row sep=-3pt},
  legend cell align=left,
]
\addplot[only marks, mark=square*, mark size=1.4pt, color=cOrd] coordinates {(1,0.97) (2,1.07) (3,1.10) (4,1.18) (5,1.26) (6,1.37) (7,1.46) (8,1.56) (9,1.64) (10,1.67) (11,1.73) (12,1.76) (13,1.89) (14,1.92) (15,2.00) (16,2.07) (17,2.13) (18,2.26) (19,2.43) (20,2.45) (21,2.48) (22,2.54) (23,2.65) (24,2.69) (25,2.70) (26,2.85) (27,2.92) (28,2.97) (29,3.22) (30,3.44) (31,3.51) (32,3.55) (33,3.70) (34,3.76) (35,3.89) (36,4.22) (37,4.25) (38,4.37) (39,4.75) (40,4.82) (41,5.23) (42,5.23) (43,5.29) (44,5.32) (45,5.33) (46,5.33) (47,5.64) (48,5.91) (49,6.38) (50,6.47) (51,6.89) (52,7.12) (53,7.27) (54,7.28) (55,7.51) (56,7.55) (57,7.71) (58,7.93) (59,8.34) (60,8.39) (61,8.42) (62,8.81) (63,9.86) (64,10.08) (65,10.19) (66,10.31) (67,10.33) (68,10.65) (69,10.78) (70,10.83) (71,11.15) (72,11.62) (73,13.27) (74,13.92) (75,14.39) (76,15.14) (77,15.38) (78,15.40) (84,20.51) (85,20.87)};
\addlegendentry{ordinary speakers (read scripts)}
\addplot[only marks, mark=star, mark size=2.4pt, color=cExp] coordinates {(79,16.96) (80,17.28) (81,17.73) (82,19.87) (83,20.27) (86,24.24) (87,24.30) (88,24.65) (89,24.68) (90,25.79) (91,25.97) (92,27.01) (93,27.61) (94,31.07) (95,32.71) (96,33.03) (97,35.09) (98,35.52) (99,36.92) (100,37.22) (101,39.02) (102,39.22) (103,41.81) (104,43.81) (105,46.84) (106,52.16)};
\addlegendentry{speech professionals (half spontaneous)}
\addplot[dashed, black!70, thin, domain=0:107, samples=2] {7.3};
\end{axis}
\end{tikzpicture}
  \caption{Sorted census CER per held-out speaker (\censusSpeakers{}
  speakers, camera A, greedy CTC), the dashed line at the median, \censusSpkMedian{}\%.}
  \label{fig:census}
\end{figure}

Fig.~\ref{fig:census} and Table~\ref{tab:census} present the census. The
population CER mean is \censusCer{}\%, from a per-speaker distribution
ranging from \censusSpkMin{} to \censusSpkMax{}\% (quartiles
\censusSpkQone{}, \censusSpkMedian{} and \censusSpkQthree{}). Gender does not explain the spread (female \censusF{}, male \censusM{}),
whereas the speaker-type annotation does. Ordinary speakers
are recognized at \censusOrdinary{}\% \ciCensusOrdinary{} and
professionals at \censusExpert{} \ciCensusExpert{} (\nOrdinarySpk{} and
\nExpertSpk{} speakers), a five-fold difference, and the professionals form
the upper tail of the distribution. The spread is a property of the
speakers rather than of the model. The per-speaker CERs of \Mone{} on the same \censusSpeakers{} speakers
correlate with those of \Mnine{} at a Spearman coefficient of
\censusSpearman{}, and \censusTopTen{} of its ten most difficult speakers
are among the ten most difficult for \Mnine{}.\footnote{Informal
inspection of the two highest- and two lowest-error speakers' recordings by
the authors agreed with the scores.}

\begin{table}[t]
  \centering
  \caption{Census by case: base CER (\%, camera A) by speaker type, speech
  mode and wording (seen = the sentence occurs in the training text), under
  greedy CTC decoding with its 95\% CI and under joint decoding, $n$
  utterances per case.}
  \label{tab:census}
  \setlength{\tabcolsep}{3pt}
  \begin{tabular}{@{}llrlr@{}}
    \toprule
    Speaker & Mode, wording & $n$ & greedy [95\% CI] & joint \\
    \midrule
    ordinary & read, seen & \nOrdReadSeen{} & \cOrdReadSeen{} \ciOrdReadSeen{} & \jOrdReadSeen{} \\
    ordinary & read, unseen & \nOrdReadUnseen{} & \cOrdReadUnseen{} \ciOrdReadUnseen{} & \jOrdReadUnseen{} \\
    professional & read, seen & \nExpReadSeen{} & \cExpReadSeen{} \ciExpReadSeen{} & \jExpReadSeen{} \\
    professional & read, unseen & \nExpReadUnseen{} & \cExpReadUnseen{} \ciExpReadUnseen{} & \jExpReadUnseen{} \\
    professional & spontaneous & \nExpSpont{} & \cExpSpont{} \ciExpSpont{} & \jExpSpont{} \\
    \bottomrule
  \end{tabular}
\end{table}

\textbf{Decomposition.} Within a speaker type and speech mode, a sentence
present in the training text is recognized O/P = \effWordingOrd{} /
\effWordingExp{} points better than one that is not. This wording effect
of the shared sentence pool is the smallest of the three. Professionals reading the same kind of script are recognized
\effDeliverySeen{} - \effDeliveryUnseen{} points worse than ordinary
speakers on matched wording, and spontaneous speech adds a further \effSpont{} points.\footnote{These
gaps are the pairwise differences of the rows in Table~\ref{tab:census},
whose intervals are pairwise disjoint. Resampling by speaker rather than
by utterance confirms all three effects. Under joint decoding every case improves by \jointCaseGainLo{} -
\jointCaseGainHi{} points (Table~\ref{tab:census},
last column), because the attention decoder completes sentences it was
trained on, the wording effect shrinking to O/P = \jEffWordingOrd{} /
\jEffWordingExp{} and the spontaneous-speech effect staying at
\jEffSpont{}.} Since most of the sentences scored by the published
protocol occur in the training text (Sec.~\ref{ssec:protocol}) and only
\expertShareSeenLo{} - \expertShareSeenHi{}\% of its utterances are
professional speech, the official score in Table~\ref{tab:official}, which follows the
protocol of \cite{park2024olkavs}, is dominated by the easiest case of
Table~\ref{tab:census}. The target application, spontaneous speech by
ordinary speakers, is a case the corpus does not contain.

\subsection{Personalization}
\label{ssec:personal}

Every one of the twelve speakers improves under both the adapter and FT-FE
with a CI excluding zero, and the gain saturates quickly. Averaged over the twelve speakers, the adapter gains \gainRtwelveSeven{}
points at 7 minutes and \gainRtwelveTwentyNine{} at 29 minutes, so more
than half of the 29-minute gain is obtained within 7 minutes, and on the
six speakers with the full budget curve it gains \gainRsixFour{} / \gainLoraRthirtytwoSeven{} /
\gainRsixFourteen{} / \gainLoraRthirtytwo{} points at 4 / 7 / 14 / 29
minutes. The gain is concentrated where
the base error is largest, and O/P speakers recover \relRecoverOrdLoR{} -
\relRecoverOrdHiR{} / \relRecoverExpLoR{} - \relRecoverExpHiR{} of their
errors. The \randomN{} randomly sampled speakers gain \randomGainR{} points
on average (\randomGainFT{} under FT-FE), with a median relative recovery
of \randomRelMedianR{}, again in proportion to the base error. We note that FT overfits within 20 steps, FT-FE gives
the largest gain for a single user (\gainFTtwelveTwentyNine{} points at 29
minutes) and serves as the reference for the attainable gain, and the
adapter is the configuration for deployment. Its rank is the decisive
hyperparameter, and rank \loraRankBest{} retains \loraShareOfFTtwelve{} of
the FT-FE gain in a \loraFileMBBest{}~MB file per user. The
LoRA-\loraRankBest{} models are used in all analyses below.

\textbf{Cross-view transfer.} Scored on every other camera available for
each speaker, the LoRA-\loraRankBest{} models adapted on camera A alone
improve \pthreeRExcluding{} of \pthreeRCells{} camera-speaker sets with a
CI excluding zero, and the gain is the same above, level with and below the
mouth plane (\pthreeRUpper{} / \pthreeRLevel{} / \pthreeRLower{} points). The
adapter therefore captures the speaker's \emph{appearance and articulation}
rather than the camera's geometry, and \emph{a single frontal recording
personalizes the model for every view} in which the device is subsequently
held.

\textbf{Effect on other speakers.} On \forgetPopSpk{} other held-out
speakers, FT-FE raises the CER by \driftFTtwelve{} points on average over
the twelve adapted models, more than the owner gains in every case, whereas
the rank-\loraRankBest{} adapter raises it by \driftRtwelve{} on average
and never by more than \driftRmax{}, \loraShareOfForgettingTwelve{} of the
FT-FE figure. In deployment the adapter is a per-user file that is detached
for other speakers, who are then unaffected.

\textbf{Visual encoder versus language decoder adaptation.}
Rank-\loraRankBest{} adapters trained on the Conformer only, the decoder
only, and both (twelve speakers, 29 minutes, joint decoding) show that the
encoder adapter alone matches the full adapter (O/P = \bOneEncOrd{} /
\bOneEncProf{} against \bOneBothOrd{} / \bOneBothProf{} points) whereas
the decoder adapter alone yields \bOneDecOrd{} / \bOneDecProf{}.
Personalization is therefore visual adaptation, and the professionals'
residual error is not a speaker-specific language problem, since adapting
the decoder to half an hour of their own speech does not reduce it.

\subsection{Camera elevation}
\label{ssec:elevation}

Every view is scored against camera A of its own group on identical
utterances with paired intervals. On the unseen-wording subset of
\Mnine{}, the three cameras above the mouth plane add \bandUpperEtwoNew{}
points on average under greedy CTC decoding, with every CI excluding
zero,\footnote{Resampling by speaker rather than by utterance confirms
it with slightly different intervals.} whereas the level cameras add
\bandLevelEtwoNew{} and the lower cameras \bandLowerEtwoNew{}. Within
camera group~2, on identical utterances, pure azimuth (C, G) costs at most
\gapEtwoVlC{} whereas the upper-center camera I, pure elevation, costs
\gapEtwoNewI{}, so the off-axis cost is due to \emph{elevation rather than
azimuth}.\footnote{The penalty is additive rather than proportional. Over
the course of training \Mnine{}, camera-A CER decreases from \trajBaseHi{}
to \trajBaseLo{}\% while the upper-camera gap remains at \trajPenHMean{} -
\trajPenIMean{} points, and a second seed reproduces it.} Training on all views keeps this penalty small. \Mone{}, trained on frontal views
only, recognizes camera A as well as \Mnine{}, but it incurs
\bandUpperEthree{} on the upper band where \Mnine{} incurs
\bandUpperEtwoNew{}, and \bandLevelEthree{} / \bandLowerEthree{} on the
level and lower bands where \Mnine{} incurs about one point and none. A
model trained on frontal views alone depends on cues that an elevated
camera removes.

\section{Discussion and limitations}
\label{sec:discussion}

The per-speaker distribution should be reported alongside the official
score, since for an individual user the distribution, not its mean, is the
relevant figure of merit.

Personalization gains are concentrated on the speakers that the base model
recognizes worst for visual reasons, mostly older ordinary speakers, and
are relatively smallest for the professionals, whose errors stem from the
wording rather than the face. The whole gain comes from an adapter on the visual encoder, whereas an
adapter on the decoder, trained on half an hour of the speaker's own
transcripts, changes nothing for any speaker. The
\effSpont{}-point penalty of spontaneous speech is thus a language-modeling
problem that a speaker's own sentences cannot solve. A count-based language
model and fine-tuning on the \bNineHours{} hours of spontaneous recordings
in the training partition do not solve it either. What is required is a language model of spontaneous Korean in general,
trained on spontaneous wording.

We cover one corpus and one language with one architecture, one
English-trained initialization and one training configuration, and
run-to-run variation is bounded by a second seed rather than by a seed
study. Elevations are qualitative, and camera identity, distance and
elevation are confounded within the array. We measure the value and the
cost of adaptation rather than rank adapters, so the adaptation methods of
prior work \cite{kim2022padding,yeo2024personalized} are not compared, and
the adapter was not swept over the layers it is attached to. Speech mode
is inferred by a rule, all personalization is frontal, and all data are
voiced studio speech, so mouthed speech may constitute a further shift
\cite{petridis2018silent}.

\section{Conclusion}
\label{sec:conclusion}

A video-only Conformer transferred from English pretraining recognizes
Korean under the OLKAVS protocol at less than half of the published error
rate,\footnote{For scale, the best published English visual-only systems report
\vallrWer{} \cite{thomas2025vallr} to \lpConformerWer{}
\cite{chang2024conformer} on LRS3, which our Korean WER on scripted studio
speech (Table~\ref{tab:official}) cannot be compared with directly.} and its held-out speakers are recognized at per-speaker CERs between 1\%
and 52\%. The corpus's own speaker annotation explains the spread. Ordinary
speakers reading scripts are recognized several times better than
professionals speaking spontaneously, with wording, delivery and
spontaneity each accounting for a measurable share of the difference, and
the official protocol samples the easiest case. Personalization on a few minutes of a
speaker's own frontal video improves high-error speakers on every
camera, and more than half of the gain is obtained within 7
minutes.
Full fine-tuning degrades other speakers by more than the owner gains,
whereas a low-rank encoder adapter retains most of the gain at a fraction
of that cost, and adapting the decoder to a speaker's own speech has no
effect. Cameras above the mouth plane add about six CER points as a constant
offset, and training on all views keeps that penalty small.

For deployment, our results suggest a simple recipe. Personalize once, frontally, use the camera level with or below the mouth,
and keep the encoder adapter as a per-user file. Spontaneous speech will require a
language model of spontaneous speech, which a user's own video does not
provide. The intended uses include an assistant that understands its user
in a crowd or on public transport, a coding assistant addressed silently
in an office, and smart glasses for people who mouth words but cannot
voice them and for people with hearing difficulty.

Future work will collect and process real-world videos from
content-sharing platforms such as YouTube, use generative models to
synthesize multiple views of frontal recordings \cite{liu2023synthvsr} and
extensions of a user's personalization video to unseen wording, train a
neural language model of spontaneous Korean, and engage professional
lipreaders for annotation and evaluation, whose performance is the ceiling
for a silent-speech interface.

\bibliographystyle{IEEEbib}
\bibliography{refs}
\end{document}